# Prediction of transcription factor binding to DNA using rule induction methods

Mikael Huss[1*] and Karin Nordström[2]

[1]Department of Numerical Analysis and Computing Science, Royal Institute of Technology, 100 44 Stockholm, Sweden
[2]Department of Cell and Organism Biology, Lund University, 223 62 Lund, Sweden; Current address: Discipline of Physiology, The University of Adelaide, Adelaide, SA 5005, Australia

*Corresponding author (phone: +46 (0)8 790 69 04)

Email addresses:
        KN: Karin.Nordstrom@adelaide.edu.au
        MH: hussm@nada.kth.se

**Supplementary information:** The paired domain dataset used in this article can be downloaded, in tab-separated text format, from http://www.nada.kth.se/~hussm/pax/pax-data.txt. The dataset in RDS format is available on request to hussm@nada.kth.se.

## Abstract

### Background

The transcription of DNA into mRNA is initiated and aided by a number of transcription factors (TFs), proteins with DNA-binding regions that attach themselves to binding sites in the DNA (transcription factor binding sites, TFBSs). As it has become apparent that both TFs and TFBSs are highly variable, tools are needed to quantify the strength of the interaction resulting from a certain TF variant binding to a certain TFBS. Ideally, one would like to have a method where any combination of TF amino acids are allowed to interact with any TFBS nucleotide, and vice versa. Rule induction algorithms might be such a method. We used a simple way to predict interactions between protein and DNA: given experimental cases from the literature where the interaction strength between two sequences has been quantified, we created training vectors for rule induction by regarding each amino acid and nucleotide position as a single feature in the example vector. The resulting interaction strength was used as the target class or value. These training vectors were then used to build a rule induction model.

### Results

We applied the rule induction method to three protein families - transcription factors from the NF-κB, the early-growth-response (EGR), and the paired domain groups – and their corresponding DNA targets. These three prediction problems increase in complexity with regard to model building, and thus gave us a good range for validation. The main focus of the study was the most complex problem: paired domain-DNA binding. For this problem, we also found sequence/binding strength correlations using measures from information theory. Prediction results were uniformly good: the rule induction approach was able to correctly order all of nine unseen examples for one NF-κB protein, achieved a correlation coefficient of 0.52 on unseen (and noisy) examples in the EGR case, and reached a classification accuracy of 69.7% for the Paired domain as evaluated by cross-validation.

### Conclusions

Rule induction methods can be useful for predicting binding strength between protein and DNA, given training examples where individual sites and classes are varied and the resulting interaction strength is quantified. Even when no knowledge about specific interactions is included, sites that have been shown to be of importance from mutagenesis and crystallography appear in the rules.

## Background

The transcription of DNA into mRNA is initiated and aided by a number of transcription factors (TFs), proteins with DNA-binding regions that attach themselves to transcription factor binding sites (TFBSs) in the DNA. As it has become apparent that both TFs and TFBSs are highly variable, tools are needed to quantify the strength of the interaction (and, subsequently, rate of DNA transcription) resulting from a certain TF variant binding to a certain TFBS. Many studies exploring the prediction of such interaction strengths have been performed (see for example [1-6]), most of which are tailored to the EGR (early growth response) zinc finger proteins. In such studies it has been common to assume independence between individual DNA-protein interactions; the binding of, say, amino acid 3 from the TF protein sequence to nucleotide 5 from the TFBS nucleotide sequence has been assumed to be independent from all other bindings. As some authors have pointed out, this independence (or "additivity") assumption is clearly unrealistic although it often gives a good approximation [7]. Ideally, one would like to have a method where any combination of TF amino acids are allowed to interact with any TFBS nucleotide, and vice versa. We believe that rule induction algorithms are such a method.

There is an additional quality of rule induction methods in the context of TF-DNA binding compared to other methods. Our initial focus was to understand the seemingly fuzzy and degenerate binding of paired domain proteins to DNA. Previously published predictive models do not seem to be readily applicable to this problem. Most of these models are either specific for EGR proteins [3, 6, 8] and thus not applicable to the paired domain. One model [5] which has been used successfully on EGR proteins, is based on data from selection experiments, but as far as we are aware, suitable data of this kind are not available for the paired domain. Two of the methods are based on scoring pairs of interacting nucleotides and amino acid residues [1, 4]. These methods could in principle be applied to the paired domain, but would at present, due to insufficient constraints on the "allowed" nucleotide-amino acid contacts, run into the problem of combinatorial explosion. Since the paired domain has 128 (or 129) amino acids that could each potentially interact with one or more of $\approx 20$ nucleotides, the number of possible configurations which would have to be scored is prohibitively high.

Rule induction seems like an ideal method for finding multi-dimensional ("many-to-one" and "one-to-many") relationships in TF-TFBS binding data. Rule induction methods use input data to construct models based on given example vectors, which contain the relevant example attributes (also called features) and the target class (in classification problems) or value (in regression problems). Rules are expressed in terms of the attribute values and can be examined and readily understood by a human user, i.e. it is always trivial to deduce the model's prediction pathway.

We used a simple way to predict interactions between TFs and TFBSs: given experimental cases from the literature where the interaction strength between two sequences has been quantified, we created training vectors for rule induction by regarding each amino acid (for TFs) and nucleotide (for TFBSs) position as a single feature in the example vector. If the interaction strength had been numerically quantified (as in the NF-κB and EGR cases), the logarithm of this value was used as a target variable for regression. If instead the interaction strength was described in terms of a few categories or classes (as in the paired domain case), the class was used as the target variable for classification. These training vectors were then used to build a rule induction model.

As it is well-known that data-driven model building is highly dependent on a suitable representation of the training data (see for instance [9]), we used two alternative representation schemes in addition to the default representation. In the default mode, amino acids and nucleotides were represented simply as letters. In the first of the alternative schemes, amino acids and nucleotides were represented in terms of their physico-chemical characteristics (see Methods). The second alternative representation described each amino acid or nucleotide with three numerical values ([10], and see Methods). The numerical representation scheme thus gives the finest granularity while the "default" is the roughest.

To evaluate the performance of a rule induction approach, we constructed models for data sets described in the literature. Two of these have been used for model building earlier – namely data for the NF-κB [11] and EGR [6] TF families. TFs in the NF-κB family co-regulate many genes and have a critical role in immunity, inflammation and apoptosis [12]. A recent study [11] first examined how well methods based on sequence profiles for NF-κB's DNA targets could predict the effects on binding specificity when the target nucleotides are varied, and then introduced an accurate method based on principal coordinates analysis. Here, only the nucleotides in the DNA target were varied, so the protein sequence was always constant within a model, although two different models corresponding to two separate NF-κB variants (p50p50 and p50p65) were constructed. This should make the NF-κB-DNA binding prediction problem somewhat easier than the EGR and paired domain problems.

Zinc finger domains are found in many eukaryotic transcription factors. Many earlier studies on prediction of protein-DNA interaction specificity have focused on zinc finger-DNA complexes, primarily for zinc finger regions from the EGR (early growth response) protein family (for example, [2, 4, 5]). In these complexes, the DNA and protein regions relevant for DNA-protein binding are known with good precision, and the identities of amino acid and nucleotide residues in the relevant sequence positions have been systematically varied in experiments. These studies have given valuable insight into the possibility of 'protein-DNA recognition codes' (which may be probabilistic in nature; see [13]),

For paired domains, no systematic experiments on this scale exist, and the regions relevant for DNA binding have not been pinpointed to the same degree as in the EGR family. Therefore no predictive paired domain-DNA binding models have been proposed before. Our main aim in initiating this study was to understand the logic of paired domain-DNA binding. Paired domain (Pax) transcription factors have fundamental roles in the development of primarily the nervous system and its associated sensory organs but also of other peripheral organ systems. Many severe human developmental defects, such as the Waardenburg syndrome [14] and Aniridia [15] are caused by paired domain mutations, and Pax defects have also been suggested to be involved in cancerogenesis (see [16, 17]). The paired domain consists of 128 amino acids organized as two helix-turn-helix motifs joined via a linker region. The crystal structure for the paired domain bound to its DNA recognition sequence has been established for two Pax proteins [18, 19]. While paired domains bind to similar consensus binding sites *in vitro*, genetically defined sites vary (e.g. [20-22]). Furthermore, the paired domain recognition site is unusually long (16-20 nucleotides) compared to other DNA-binding proteins.

Due to the nature of available data, the difficulty of predicting interaction strength increases in the three example families. In the first data set (NF-κB [11]) only DNA target sequences vary. In the second (EGR [6]) both amino acids and nucleotides vary, but only in a limited number of positions. After showing that rule induction can give good results on previously studied datasets, without having been tuned to incorporate any domain specific information about protein-DNA contacts, we attacked the difficult problem of paired domain-DNA binding, which is known to bind to DNA targets with a degenerate and fuzzy binding code [23]. With the method described here, more subtle interactions that are not immediately deduced from crystallography data can be identified. We also evaluated whether including information gain profiles of individual sites would improve the prediction quality, and found this not to be the case. Using the entire problem space, we found that paired domain-DNA interaction can be predicted with an accuracy ($\approx 70\%$) which is far from perfect, but statistically significant compared to a random guess ($\approx 43\%$). More importantly to the molecular biologist, we found a number of highly predictive and biologically understandable rules that can be used to predict and understand paired domain-DNA binding.

# Results

### Applying rule induction to NF-κB-DNA binding data

We started by building regression models for NF-κB TF data (from [11]). This problem should be the easiest to solve, since only the DNA target sequences, and not the TFs varied (although two alternative TFs were examined). We used the same training data as in the original study, which consisted of 52 examples. For these example cases, there were ten features corresponding to nucleotide position identities plus the logarithm of the interaction strength with either the p50p50 or the p50p65 variant of NF-κB as the target variable. We tried all three representations of the nucleotides and amino acids.

We built regression models for both p50p50 and p50p65 and assessed the models' accuracy by leave-one-out cross-validation and calculating the correlation coefficient between the model's prediction and the logarithm of the measured interaction strength. (The correlation coefficient varies between $-1$ and $1$, the latter representing perfect correlation and the former perfect anti-correlation. The value 0 represents a lack of correlation.) For p50p50, the two best models, which were both based on bagging and recursive partitioning (using two different types of data representation), had correlation coefficients of 0.77 and 0.78 (see Table 1). Results in the original study ranged from 0.74 to 0.90 (see Table 1 in [11]) .

For p50p65, the same model settings (recursive partitioning and bagging with default or numerical representation) gave the best results (0.61 and 0.64 respectively). The original study (see Table 1 in [11]) found correlation coefficients from 0.66 to 0.82.

After thus having evaluated the fit of the p50p50 and p50p65 models to training data, we also evaluated their performance on the same test set as in the original study [11], consisting of 9 new DNA targets which were not used for building the models. For these, no binding strength values were available, but their relative rank with regard to other DNA targets were known. Therefore we tested our best models to see if they could correctly order the 9 unknown targets from lowest to highest affinity. We measured the error by counting how many stepwise movements of symbols in the string that represents the ordering that have to be made in order to reach the true ordering. (We assume that the true ordering is represented by the string '123456789'. This ordering thus has error 0, while the worst possible ordering '987654321' has error 36.) For the p50p50 case, it turns out that according to this error definition the best models have test errors of 2 and 3, respectively. (For illustration, in the first case, the predicted order is 123**465879** instead of 123456789. Moving each of the symbols '6' and '8' one step to the right yields the true ordering. ) For the p50p65 case, the best model manages to perfectly order the DNA targets (test error=0) while the second best model has a test error of 4.

Noting that the performance of our method, although worse than more specialized approaches, is satisfactory on this problem (Table 1) we moved on to EGR zinc finger proteins, a highly modeled family of TFs.

**Applying rule induction to zinc finger-DNA binding data**

We tested our method on a previously analyzed dataset [6]. In this study, a microarray approach was used to generate a comprehensive dataset, where the binding of wild-type zinc fingers and four mutant zinc finger variants to all their possible tri-nucleotide targets was quantified (see [6] for more details on zinc fingers and this data set in particular). Because of the systematic design of the experiments used to arrive at this dataset, we were able to use the binding measurements (after taking logarithms) as continuous target values for regression. We used RDS to construct a range of models (using recursive partitioning, covering and bagging) from this data set, and tested the resulting models' performance on independently reported zinc finger-DNA binding experiments [24]. The training examples were vectors consisting of ten amino acids followed by three nucleotides involved in direct binding, and the target variable (the natural logarithm of $K_d$). We tried all three representations of the nucleotides and amino acids.

We evaluated the performance using the correlation coefficient. The models were able to learn the training data set quite well, with correlation coefficients of up to 0.84 (Table 2), as assessed by leave-one-out cross-validation. Upon testing on the independent cases, the best model (covering, numerical representation) gave a correlation coefficient of 0.52. We find this performance to be satisfactory, considering that (i) we have not included any problem-specific information about individual nucleotide-amino acid contacts, and (ii) the test cases come from experiments performed in a different lab with different experimental conditions. Indeed, one of the experiments in the test data [24] was also performed in the training data set [6] and the reported binding strength differs by a considerable margin between the two. (In the training data set, the protein-DNA combination in question has a $K_d$ of $\approx 0.38$, while in the test set, it has $K_d \approx 0.15$.)

Interestingly, we found that the models based on recursive partitioning and/or bagging resulted in overfitting on this problem. The covering models, which yield more compact hypotheses with fewer rules, were able to generalize better (Table 2). We hypothesize that the sampling of amino acid sequences was insufficient (there were only five zinc finger variants) to build a detailed model.

As we obtained good performance level on this microarray-based study with both amino acid and nucleotide sequences varied, and we had thus shown that rule induction can give good results on previously studied datasets, without having been tuned to incorporate any domain specific information about protein-DNA contacts, we attacked the more difficult problem of paired domain-DNA binding.

**Application of an information gain measure on paired domain-DNA binding data**

Since the paired domain was expected to be harder to model, we wanted to extract sites with the highest information gain for fine-tuning of the training data. We collected 598 binding experiments from the literature and assigned the interaction strength in each experiment to one of three classes (++: strong interaction, +: weak interaction, -: no interaction). Using the full set, we calculated information gain profiles quantifying the usefulness of each single sequence position for predicting binding strength (Figure 1a, b). For the purposes of these calculations, we considered only the identity of each nucleotide or amino acid, with no special encoding of their physico-chemical character or other features. Specifically, we used the information gain ratio instead of the standard information gain measure, which favors attributes that can take on a large number of values [25]. Many amino acid positions that we found to show substantial information content have previously been described to be associated with binding specificity. These include position 17 and 48 (Figure 1a) that have been shown to be naturally occurring Pax3 mutations in families with the Waardenburg syndrome [26], and residues in areas where many residues contact the DNA directly [18, 19], such as positions 70 and 71. These two have also been shown to be responsible for the lack of binding in *Acropora millepora* PaxD [27]. Other results were somewhat surprising and included amino acid residues 84, 88, and 109, located in regions of low or no direct DNA contact. One can hypothesize that they have a crucial role in maintaining the protein 3D backbone.

We also calculated information gain ratio profiles quantifying the usefulness of all combinations of nucleotide and amino acid pairs for predicting the binding strength (Figure 1c, for the 20 most informative combinations). A few sites appear several times in this list. These include amino acid residues 101 and 104, two sites with low individual information gain, but which are fairly conserved across wildtype paired domains [28]. Nucleotides bound by the N-terminal half of the paired domain (nucleotides 10-21 [18]) provide more information for determining binding specificity (Figure 1c), which is supported by the fact that the C-terminal is dispensable for DNA-binding by some paired domains [19, 29].

Interestingly, these preliminary studies were of little value when it came to building global prediction models (see below). The single positions with high information gain values did not necessarily appear in any reliable classification rules. Apparently, combinations of several amino acid and nucleotide sequence positions are necessary to determine binding specificity; knowing one or two positions is not enough.

**Applying rule induction methods on paired domain-DNA binding data**

Using the same 598 case data set as above, we applied rule induction to the paired domain. The best prediction results were obtained using the numerical representation (see Methods) and a bagging algorithm on top of recursive partitioning (Table 3). The predictive accuracy was estimated using 5-fold cross-validation, which means that results from five runs, in each of which 80% of the data was used for building the predictive model and 20% was used purely for testing the resulting model's accuracy, were averaged. The 20% of the examples used for "blind testing" in each of the 5 runs were cycled so that in the end, all examples had been used for testing. Bagging with recursive partitioning yielded an accuracy of 69.7%, with a default accuracy of 42.7% (Table 3).

In addition, we evaluated a feature selection strategy, where only the 15 amino acid attributes with the highest information gain ratios according to our preliminary studies were used in building the prediction model. This model performed much worse than the models built from all amino acid attributes (not shown). The explanation for this is probably that individual sequence positions are not informative enough; the correlations between sequences and binding strengths occur at the level of combinations of nucleotide and amino acid positions. This can also be seen by inspecting the classification rules from the best models.

The vast majority of classification rules (Figure 2) include both amino acid and nucleotide attributes, underlining the fact that neither the paired domain nor the target DNA site can in themselves guarantee good binding; it is the relation between the two that determines the binding strength. Even if there are not enough 'good' rules to cover the whole problem space, those rules that we found to have high accuracy are highly useful where applicable.

## Conclusions

We have shown that rule induction can be applied to TF-DNA binding specificity prediction. No prior knowledge about DNA binding has been incorporated and all sequence positions have been considered equally. The results therefore provide an unbiased model of protein-DNA interactions. The satisfactory performance of the method on data from different TF families shows its strength and general validity. Our approach is reasonably successful on previously analyzed TF families, like NF-κB and EGR, but we mainly want to introduce it as a powerful tool in more complex cases, such as paired domain-DNA interaction prediction. Our method, applied to this difficult problem, shows a fair if unspectacular prediction performance, but importantly, it identifies a number of classification rules of high predictive accuracy. These rules are expressed in terms of both amino acids and nucleotides, and may help in identifying true physical interactions. Our study also pointed out two interesting details. Firstly, in the paired domain case, data pre-processing by feature selection using information gain (which measured correlations between sequence positions or pairs of positions and the binding strength) was useless for building the prediction models. This suggests that paired domain-DNA binding specificity is determined by higher-level, complex interaction patterns where the identities of individual residues are relatively unimportant in themselves: the right combination is what matters. Secondly, we found that data representation was fairly unimportant in this problem domain. The relatively sophisticated numerical representation did, on the whole, perform best, but only by a small margin compared to the rough default method.

## Methods

### Binding data

*NF-κB binding data.* We used NF-κB DNA-targets (supplementary information from [11]), which contains 52 DNA sequences. Each sequence contains 10 nucleotides. Scores from two replicates, along with the geometric mean of these scores, are reported for the strength of the interaction of each DNA target with two NF-κB variants: p50p50 and p50p65. We constructed two datasets, one for each NF-κB variant. Each example vector contained 10 features corresponding to the 10 nucleotides, followed by the natural logarithm of the geometric mean of the interaction strength value with respect to p50p50 or p50p65, dependent on the data set.

*EGR binding data.* We used a previously studied comprehensive dataset (see supplementary material in [6]), where the binding of wild-type zinc fingers and four mutant zinc finger variants to all their possible trinucleotide targets was quantified. Our training vectors consisted of the amino acids of zinc finger 2 involved in binding (10 in number), followed by the interacting nucleotides (3 in number) and the target variable (the natural logarithm of the $K_d$ value of the interaction; as there were several replicates of each experiment, we used the average $K_d$ value in each case). Finally we tested the resulting models' performance on independently reported zinc finger-DNA binding experiments from the literature [24]. The examples found in that paper were manually converted into the same format with 10 amino acids, 3 nucleotides and (the natural logarithm of) one $K_d$ value.

*Paired domain binding data.* For paired domains we had to retrieve data from a large range of publications and treat them in a consistent way to build our example vectors.

Paired domains can be either 128 or 129 amino acids long, as paired domains of the Pax3/7 class have an extra amino acid inserted at position 75. We thus used 129 attributes for the amino acid positions, and set position 75 to a sequence gap in all the paired domains with 128 amino acids. We elected to use 21 attributes (corresponding to 21 nucleotide positions) in the DNA sequence targeted by the paired domain. The specific nucleotide positions used in each case were obtained by making alignments of the DNA sequences using a few early publications as primary guidelines (mainly [21, 23, 30, 31]). We thereafter assumed that the corresponding nucleotide positions were comparable as far as amino acid - nucleotide interactions were concerned. Due to this way of presenting and analyzing the data set, DNA sequences that were difficult to align convincingly had to be left out from the remaining analysis. Since the binding strengths were reported in a wide variety of ways in the articles, from numerical values to pictures of spots on a gel, the assignment of classes to examples was a difficult task. The dataset is therefore, in machine learning jargon, noisy. As an additional difficulty, the data set is not evenly sampled: the available binding data in many cases uses different DNA sequences for each paired domain.

Binding data was retrieved from the following references: [18, 20, 21, 23, 26, 27, 30-57]. We used binding data for paired domains only when these could be analyzed without interference from homeodomains or other parts of the Pax sequence. The complete set consisting of 598 binding experiments used to build and evaluate the models. 158 DNA-sites were aligned with ClustalW [58], adjusted by eye, and trimmed down to 21 nucleotides. The sites were adjusted using a range of sites tested ([23] as a guide, using their nucleotides 4-24, as well as [21, 30, 31]). No gaps were allowed in the nucleotide alignment. Nucleotide sequences that could not be convincingly aligned with the other sites, such as 5aCON [38], were excluded from the remaining analyses. 103 complete 128-129 amino acid paired domains (wildtype and those altered by mutagenesis) were aligned using ClustalW. Gaps were allowed between positions 74 and 75, where wildtype Pax3 and Pax7 have an additional residue, but at no other positions.

Binding strengths were extracted from the publications and given three possible denominations: '-' for no binding, '+' for weak or ambiguous binding, and '++' for definite binding. Since much published binding data is unquantified, we used comparisons between known sites in different experiments to determine the level of binding, using [23] as a primary guide. For several paired domain–DNA binding site combinations, multiple references were found, confirming our initial judgments. Ambiguities between publications led to the exclusion of the data point(s) in question.

**Sequence descriptors**

We initially modeled each sequence with one attribute for each of its nucleotides, allowing as a value either a letter corresponding to each of the four naturally occurring nucleotides, or 'N' for unknown, followed by the attributes corresponding to the amino acid residues, allowing as a value a letter corresponding to either one of the 20 naturally occurring amino acids or a hyphen ('-') for a gap. This initial coding was identical for each of the three representation schemes. In the default scheme the initial representation was not modified. In the "numerical", representation scheme, each letter was expanded into three numerical values corresponding to three biophysical descriptor scales of amino acids and nucleotides, respectively [10, 59]. These descriptors were generated by compressing a large number of previously used descriptors using principal component analysis and the partial least squares algorithm [10] to yield more compact, compound descriptors. The nucleotide descriptors are derived from 24 different physico-chemical properties and can be said to represent size, electronic properties and hydrophobic properties. The amino acid descriptors are derived from 26 physico-chemical properties, and the three descriptors we use roughly correspond to, respectively, a hydrophobicity measure, a steric bulk measure, and a polarity measure [10, 59]. In the enriched representation scheme, only the letters corresponding to amino acid positions were expanded. It described amino acids in terms of polarity (polar or nonpolar), hydrophobicity (hydrophobic or non-hydrophobic), charge (charged or not) and "special characteristics" (whether the amino acid has "special properties" like the –SH group in cysteine and the ring structure in proline; apart from these, glycine is also included among the "special" amino acids). Nucleotides were described as purines or pyrimidines.

**Information gain**

The information gain (*IG*) for an attribute or a *feature* (here denoted as *f*) with respect to the set of classes (here denoted as *X*) was calculated as:

$$IG(f,X) = H(X) - H(X|f)$$

where *H(X)* is the entropy of *X* :

$$H(X) = -\sum_{x \in X} P(x) \ln P(x)$$

and *H(X|f)* is the entropy of *X* when the value *v* of feature *f* is known:

$$H(X|f) = -\sum_{v} P(v) \sum_{x \in X} P(x|v) \ln P(x|v).$$

*P(x)* is the probability that an example belongs to class *x*, *P(v)* is the probability that feature *f* has value *v*, and *P(x|v)* is the probability that an example belongs to class *x* given that feature *f* has value *v*.

However, the information gain tends to favor attributes with many possible values. Thus for example, an attribute corresponding to a highly variable amino acid position might get a high information gain value even if the variations are not strongly correlated with the class. Therefore, we used the information gain ratio (*IG-ratio*), which is the information gain divided by a correction term that describes information needed to determine the value of the attribute:

$$IG\text{-}ratio\ (f,X) = IG(f,X) / ( -\sum_v P(v)\ ln\ P(v)\ ).$$

Above, the observed nucleotides and amino acids at specific sequence positions were considered as the outcomes of random variables with alphabets $\Omega_n$ and $\Omega_a$ comprising, respectively, letters corresponding to the four nucleotides (A, G, T, C) and 'N' for unknown, and letters corresponding to 20 natural amino acids and '-' for a gap. Similarly, the binding strength was interpreted as the outcome of a random variable $X$ with possible outcomes in $\Omega_c = (-, +, ++)$.

**Rule induction**

We used the rule induction package Rule Discovery System (RDS [60]) to build models from the example cases. RDS allows the user to create classification rules either in the form of decision trees (using *recursive partitioning*) or lists of potentially overlapping rules (using *covering*). Recursive partitioning gives rise to decision trees, which are appropriate for the task at hand, as they handle discrete data, evaluate information context-specifically, and represent extracted knowledge in a comprehensible manner. Covering algorithms have the additional advantage of often yielding more compact hypotheses than decision trees [61].

In addition to these methods, RDS also supports ensemble methods such as *bagging,* which works by generating a set of hypotheses by repeated subsampling of the data. At prediction time, the predictions of different hypotheses are combined to yield a single prediction. RDS furthermore allows the user to specify background knowledge about the problem domain in PROLOG format.

# Authors' contributions

KN provided the initial idea to the project, assembled the paired domain data set and wrote part of the article. MH performed the rule induction and information gain experiments and wrote part of the manuscript.

# Acknowledgements

We thank Per Lidén at Compumine for valuable discussions and Joel Westerberg for computer assistance.

# Figure Legends

### Figure 1 – Paired domain information gain profiles

In a) a consensus paired domain with the conserved residues (invariant in most wildtype proteins [62]) shown underlined. Sites that vary too much to reach a meaningful consensus are shown with a period ('.'). DNA contact points of Pax6 and Paired [18, 19] are shown under the sequence with 'x'. Between position 74 and 75 (where Pax3 and Pax7 have an additional residue) a gap in the alignment is shown with a hyphen ('-'). The information gain ratios for the 15 most informative sites are shown as histobars directly under the consensus sequence. The height of the bar is proportional to the information gain ratio of the corresponding site. Note that many of the positions with high information peaks are located close to residues that make DNA contacts. Panel b) shows a nucleotide alignment of a subset of the DNA sites analyzed, together with the information gain profile for the ten most informative nucleotide sites under the alignment. The sites are named as in their primary reference. The N-terminal end binds to site 10-21, the C-terminal to 1-7, and the linker to 8-15. In c) the top 20 combinations of sites are shown in the order of the information they provide. Column 1 shows nucleotide positions numbered 1-21 as in the alignment in panel b); column 2 shows the amino acid residues numbered 1-128 as in panel a) above; and 3) finally shows the information that doublet provides.

### Figure 2 – Paired domain decision rule examples

The decision rules were extracted with high prediction accuracy, and were generated using three different data representations in RDS. Note that all rules use both nucleotides (nt) and amino acid (aa) residues to predict binding strength. The rules consist of logical statements involving the nucleotide and amino acid positions. In the default representation, only the identity of each amino acid or nucleotide is considered. In the enriched representation, each amino acid is also described with attributes relating to hydrophobicity, charge, aromaticity, and special characteristics. Thus, rules generated using this representation can contain logical relations of the form "amino acid 2 is hydrophobic", in addition to rules involving residue identities. In the numerical representation, nucleotides and amino acid residues are encoded by numerical biophysical descriptor values, which are then used to generate rules. The three nucleotide descriptors roughly represent size, electronic properties and hydrophobic properties, and the three amino acid descriptors roughly correspond to a hydrophobicity measure, a steric bulk measure, and a polarity measure [10, 59]. In this case, rules are not of the form "X is Y" or "X is not Y", but of the form "X is larger than Y" (or "X is larger than or equal to Y") or "X is smaller than Y" (or "X is smaller than or equal to Y"). The Predictions line shows the predicted binding class for examples matching the rule shown. The Reliability line lists the performance of the rule on example cases in the training set (which consists of a randomly selected 80% of the examples) and the test set (which consists of the remaining 20%).

# Tables

**Table 1. Prediction performance on NF-κB data using a selection of methods and data representations.**

| NF-κB variant | Method | Representation[1] | CC (tr)[2] | test error[3] |
|---|---|---|---|---|
| P50p50 | Bagging with rec.part. | Default | 0.77 | 2 |
| P50p50 | Bagging with rec. part. | Numerical | 0.78 | 3 |
| P50p65 | Bagging with rec. part. | Default | 0.61 | 4 |
| P50p65 | Bagging with rec. part. | Numerical | 0.64 | 0 |

---

[1] Refers to default or numerical feature representation.

[2] Correlation coefficient, estimated by leave-one-out cross-validation on the training data.

[3] The test error measures how far from the true ordering the model's ordering of the 9 test examples is. The ordering can be expressed as a string, where '123456789' would correspond to the true ordering (test error = 0). The test error indicates how many stepwise movements of the characters in the string that would have to be applied until the true ordering has been reached. For example, the string '152346789' has test error = 3 since the number 5 must be moved 3 steps to the left to give the right ordering. The worst possible ordering, '987654321', has test error = 36.

**Table 2. Prediction performance on zinc fingers using a selection of methods and data representations (320 data points in training data, 14 data points in test data). Note that the especially the recursive partitioning-based methods suffer from heavy overfitting; the training set results are much better than the test set results.**

| Method | Representation[4] | CC (tr)[5] | CC (test)[6] |
|---|---|---|---|
| Covering | Numerical | 0.68 | 0.52 |
| Covering | Default | 0.69 | 0.36 |
| Recursive partitioning | Default | 0.74 | 0.35 |
| Bagging with rec. part. | Numerical | 0.84 | 0.14 |

---

[4] Refers to default or numerical feature representation.

[5] Correlation coefficient, estimated by leave-one-out cross-validation on the training data.

[6] Correlation coefficient, estimated by testing on 14 independent examples collected from an independent study [24].

**Table 3. Prediction performance on the Paired domain using different methods and data representations (598 data points). The performance was evaluated by 5-fold cross-validation.**

| Method | Default[7] | Enriched[8] | Numerical values[9] |
|---|---|---|---|
| Recursive partitioning | 54.7% | 54.7% | 53.0% |
| Covering | 57.7% | 59.4% | 55.0% |
| Bagging with rec. part. | 65.0% | 62.9% | 69.7% |
| Bagging with covering | 60.7% | 61.0% | 59.2% |

---

[7] Default feature representation.

[8] Prediction accuracy using the enriched feature representation.

[9] Prediction accuracy when amino acid and nucleotide residues were assigned numerical values.

```
1        10        20        30        40        50        60
G.GGVNQLGGVFVNGRPLPD.IR.RIVELAH.GVRPCDISRQLRVSHGCVSKILGRYYETGSIR
x    xx    x xxxxx    x          xxx   x   xxxxx xx    x
          □  □□   □           □  □     □□

        70        80        90       100       110       120
PG.IGGSKPK-VATP.VV.KI.EYKRENPTIFAWEIRDRLL.EGVCD..NVPSVSSINRILRNK.
xxxxxxx xx xx               xxx        □       □        □
     □□          □    □
```

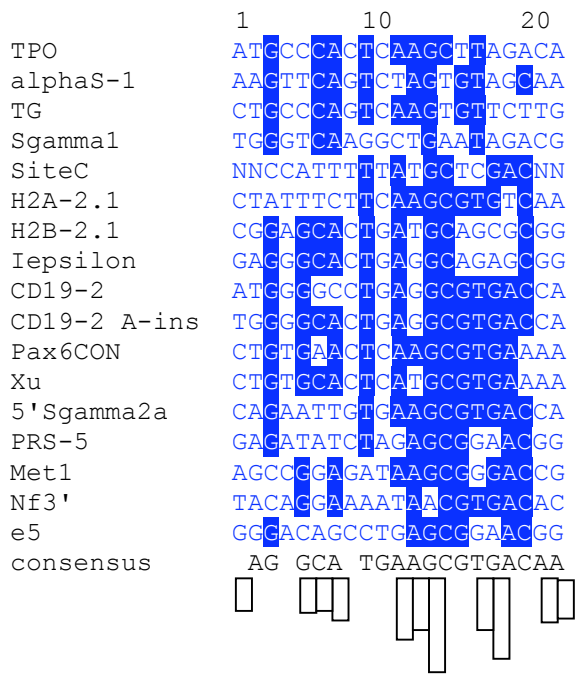

| Nucleotide[1] | Amino acid[2] | Inf. Gain ratio[3] |
| --- | --- | --- |
| 13 | 104 | 0.0906 |
| 18 | 128 | 0.0884 |
| 13 | 128 | 0.0848 |
| 13 | 101 | 0.0832 |
| 13 | 70 | 0.0829 |
| 13 | 71 | 0.0829 |
| 15 | 128 | 0.0816 |
| 13 | 37 | 0.0809 |
| 13 | 17 | 0.0798 |
| 17 | 101 | 0.0797 |
| 13 | 117 | 0.0793 |
| 6 | 128 | 0.0793 |
| 18 | 21 | 0.0791 |
| 13 | 80 | 0.0781 |
| 13 | 55 | 0.0781 |
| 14 | 128 | 0.0763 |
| 17 | 128 | 0.0761 |
| 16 | 128 | 0.0760 |
| 18 | 79 | 0.0756 |
| 17 | 104 | 0.0754 |

**Example rule 1.**
Generated by the default representation.

    nt 1 **is** T
    aa 55 **is not** G
    aa 70 **is not** G

---

    Prediction: - (no binding)
    Reliability: 5/5 (training set), 2/2 (test set)

**Example rule 2.**
Generated by the default representation.

    nt 17 **is** G
    aa 37 **is not** C
    aa 75 **is** Q
    aa 81 **is not** I

---

    Prediction: ++ (strong binding)
    Reliability: 10/10 (training set), 2/2 (test set)

**Example rule 3.**
Generated by the enriched representation.

    nt 13 **is** G
    nt 14 **is** C
    nt 15 **is** G
    nt 19 **is** C
    aa 55 **is** polar

---

    Prediction: + (weak or ambiguous binding)
    Reliability: 11/11 (training set), 4/4 (test set)

**Example rule 4.**
Generated by the numerical representation.

    nt 1 **has steric bulk descriptor value less than or equal to** -0.8203
    aa 70 **has hydrophobicity descriptor value less than or equal to** $-1.115$
    aa 82 **has hydrophobicity descriptor value larger than** 2.17

---

    Prediction: - (no binding)
    Reliability: 11/11 (training set), 3/3 (test set)